\documentclass[12pt,a4paper]{article}

\usepackage{times}
\usepackage{natbib}
\usepackage{aas_macros}

\begin{document}
\renewcommand{\thefootnote}{\arabic{footnote}}
\title{Solar wind origin of terrestrial water}
\author{Hans Merkl (1)  and Markus Fr{\"a}nz (2)\\
\small (1) {Am Kirchsteig 4, 92637 Weiden, Germany}\\
\small (2) {Max-Planck-Institute for Solar System Research, 37191 Katlenburg-Lindau, Germany}}
\date{2011/01/28}
\maketitle

\abstract{The origin of the Earth water reserves during the evolution of the planet is one
of the big miracles in geophysics. 
Common explanations are storage of water in the Earth mantle at a time when the crust
had not yet formed and depositing of water by comets during the time of late heavy bombardement.
Both explanations have different problems - especially when comparing with the evolution
of Mars and Venus. Here we discuss the possible role of hydrogen collected from the solar wind
by the early Earth magnetosphere. While the water production by solar wind capture 
is very small today it may have been significant during the first billion years after planetary formation
because solar wind was much stronger at that time and Earth magnetospheric configuration
may have been different. We estimate that the contribution of solar wind hydrogen to the Earth
water reserves can be up to 10\%  when we assume a that the Earth dipole acted as a collector
and early solar wind was 1000 times stronger than today. We can not even exclude that
solar wind hydrogen was the main contributor to Earth water reserves}

\section{Introduction}
Water is one of the essential ingredients for the evolution of life on a planet.
Among the terrestrial planets Earth is unique in its amount of water - stored
in its oceans. Still no commonly accepted theory exists on the origin of this water
(see for example \cite{williams:2007}).      According to current models of the
formation of our planet Earth formed from the solar nebula about 4.54 billion
years ago and its surface remained in a hot fluid magma state for the next 400 million
years. In a geological explanation of the origin of water this early magma contained
enough hydrogen and oxygen to cause an outgassing of water when the crust
of Earth cooled down \cite[]{williams:2007}.  An astronomical explanation invokes
the import of water from comets during the time of late heavy bombardment
about 4.1 to 3.8 billion years ago. Both explanations do not agree with the recent
discovery of zirconium minerals indicating fluid water on the Earth surface about 4.4 billion
years ago \cite[]{wilde:2001}. Another problem of both explanations is that - if these
explanations hold - both Mars and Venus should harbor amounts of water comparable to Earth which they obviously
do not do at present time. A common explanation for this is that Mars and Venus 
had a similar atmospheric composition after the formation of the planets but
have suffered from atmospheric erosion by solar radiation more than Earth \cite[]{lammer:2008}. 
One main reason for the difference in erosion in this explanation is the protection
of Earth by its magnetic field. While in the models of atmospheric erosion
the ablation of unprotected atmospheres by solar wind impact is regarded as
an important factor it has so far not been considered how strong
the import of solar wind hydrogen into a large magnetosphere can be.
In this short article we will elucidate this factor on the evolution
of terrestrial water reserves.

\section{Intrusion of protons into the magnetosphere}
The effect of a magnetosphere on the solar wind flow is manifold. In the configuration
of the Earth magnetosphere as we see it today the magnetic poles are rather
close to the rotation axis and do never point parallel to the direction of the
solar wind. In this case the magnetosphere acts as a shield but part of the
solar wind flux is deflected towards the magnetic poles where particles
can penetrate to low altitudes along the field lines. Measurements of the 
penetrating proton flux have been collected by \cite{hardy:1989} who report mean fluxes
of the order of 10$^{24}$ protons/s penetrating over one pole for current
mean solar wind conditions with a variation of one order of magnitude.
It is important to note that the total proton flux is linearly dependent on
the solar wind intensity and the size of the polar cusp directed towards
the sun which can be much larger than a planetary diameter. Another
effect which may be important is the storage of solar wind protons 
in the magnetospheric tail. Part of these protons can also
penetrate to low altitudes in magnetospheric storms.
\cite{hardy:1989} already realized that these protons can be a significant
source of hydrogen at altitudes above 300km. If we now assume that
an early planetary atmosphere was rich in carbon-dioxide and oxygen
from photo-dissociation the penetrating hydrogen may produce a 
significant amount of water.

For Venus this collection of protons through a large magnetic cusp never happened
since Venus has no internal magnetic field. For Mars we know that the planet had 
an early magnetic field that  probably lasted for several hundred million years.
In our model this time span may have been long enough to collect the amount
of water which we still see today in the Martian water-ice reserves.

\section{Early solar wind intensity}
The current solar wind has a mean flow velocity of about 400km/s and an
intensity of about 1.0/cm$^3$ at Earth corresponding to a flux intensity of  
$F=n\times v = 4\times10^7/$cm$^2$s. The amount of protons impinging
on the total magnotospheric cross-section with a diameter of about 10 Earth radii
is $F\times A = 5\times 10^{27}$/s, where $A=100\pi R_E^2=1.3\times10^{20}$cm$^2$.
Taking the results of \cite{hardy:1989} this means that at current conditions
only about 0.1\% of the total impinging flux reaches the Earth atmosphere.
Taking the proton mass of $m_p=1.7\times10^{-27}$kg the current proton
mass import is only about $M=0.001\times m_p\times  F\times A = 0.01$kg/s.
If any two protons react with free oxygen this corresponds to about 0.08kg/s of water
or about 2500 tons/year. The total water content of the Earth atmosphere 
is about $2\times10^{13}$tons, of the oceans $1.4\times10^{18}$tons.
This means at current solar wind level and magnetospheric configuration
the solar proton import can just refill the atmospheric water content 
over a time span of the age of the solar system. 
Observations of sun-like stars show that the particle flux from early sun
may have been up to 1000 times stronger than the current solar wind
for a period of 500-1000 million years after the formation of the solar system
(see for example the review by \cite{lammer:2008}). 

\section{Early Earth magnetic field}
While many models of the early Earth predict a lower magnetic field intensity
of the Earth, recent observations show that a significant Earth magnetic field existed at
least 3.5 billion years old \cite[]{tarduno:2010}. If we now assume that
the Earth magnetic field configuration was multipolar or its cusp directed
more towards the sun the magnetosphere may have acted as a solar
wind collector and not as a shield. Taking a cusp cross section of 1 Earth radius
in such a geometry would result in a proton flux of $F\times A = 5\times 10^{25}$/s
for current solar wind intensity and $F\times A = 5\times 10^{28}$/s for strong early
solar wind corresponding to $2\times10^{8}$tons of water  per year. Over a time span of 1 billion years
the solar proton influx then may have contributed up to 10\% of the
total Earth water reserves. If we assume that the early Earth magnetic field
was even stronger and its cusp directed towards the Sun one can even reach a higher
estimate.   Since this is only an order of magnitude
estimate we can not even exclude that solar wind hydrogen was
the main contributor to the Earth water reserves.  

\section{Deuterium content of planetary waters}
The relative deuterium content of planetary water reserves
is usually taken as indicator for the amount of lighter hydrogen
atoms which have been escaping from the atmospheres over
the age of the solar system (see for example \cite[]{greenwood:2008} for recent Martian deuterium observations). 
The current deuterium/hydrogen
ratios are 0.015\% for Earth, 0.08\% for Mars, and 2.0\% for Venus, but
0.002\% for the solar wind and 0.031\% for the few comets where
measurements have been possible \cite[]{lammer:2008}. If the original deuterium ratio of the Earth
water reserves was similar to the cometary one the influx of solar wind
protons could explain the lower ratio observed on Earth today.

\section{Discussion and conclusions}
We have shown that the influx of solar wind protons into the Earth atmosphere
may be a significant source of water during the early evolution of the solar system
if the Earth magnetosphere acted as a solar wind collector.
While for planets without a magnetosphere, like Mars and Venus, the escape of hydrogen and oxygen is
usually regarded  as higher than the proton influx, this is not so clear for Earth.
Only recently \cite{engwall:2009} reported that the current hydrogen outflow
may be 10$^{26}$ protons/s - significantly higher than the inflow observed
by \cite{hardy:1989}. But this must not hold for early solar system conditions.
Also for other solar system bodies with magnetospheres, for example moons of the outer planets, the collection of ions
through magnetospheric cusps may be an important source of atmospheric particles.

\small
\bibliographystyle{agu}
\bibliography{mars}

\begin{thebibliography}{7}
\providecommand{\natexlab}[1]{#1}
\expandafter\ifx\csname urlstyle\endcsname\relax
  \providecommand{\doi}[1]{doi:\discretionary{}{}{}#1}\else
  \providecommand{\doi}{doi:\discretionary{}{}{}\begingroup
  \urlstyle{rm}\Url}\fi

\bibitem[{\textit{{Engwall} et~al.}(2009)\textit{{Engwall}, {Eriksson},
  {Cully}, {Andr{\'e}}, {Torbert}, and {Vaith}}}]{engwall:2009}
{Engwall}, E., A.~I. {Eriksson}, C.~M. {Cully}, M.~{Andr{\'e}}, R.~{Torbert},
  and H.~{Vaith}, {Earth's ionospheric outflow dominated by hidden cold
  plasma}, \textit{Nature Geoscience}, \textit{2}, 24--27,
  \doi{10.1038/ngeo387}, 2009.

\bibitem[{\textit{{Greenwood} et~al.}(2008)\textit{{Greenwood}, {Itoh},
  {Sakamoto}, {Vicenzi}, and {Yurimoto}}}]{greenwood:2008}
{Greenwood}, J.~P., S.~{Itoh}, N.~{Sakamoto}, E.~P. {Vicenzi}, and
  H.~{Yurimoto}, {Hydrogen isotope evidence for loss of water from Mars through
  time}, \textit{\grl}, \textit{35}, 5203--+, \doi{10.1029/2007GL032721}, 2008.

\bibitem[{\textit{{Hardy} et~al.}(1989)\textit{{Hardy}, {Gussenhoven}, and
  {Brautigam}}}]{hardy:1989}
{Hardy}, D.~A., M.~S. {Gussenhoven}, and D.~{Brautigam}, {A statistical model
  of auroral ion precipitation}, \textit{\jgr}, \textit{94}, 370--392,
  \doi{10.1029/JA094iA01p00370}, 1989.

\bibitem[{\textit{{Lammer} et~al.}(2008)\textit{{Lammer}, {Kasting},
  {Chassefi{\`e}re}, {Johnson}, {Kulikov}, and {Tian}}}]{lammer:2008}
{Lammer}, H., J.~F. {Kasting}, E.~{Chassefi{\`e}re}, R.~E. {Johnson}, Y.~N.
  {Kulikov}, and F.~{Tian}, {Atmospheric Escape and Evolution of Terrestrial
  Planets and Satellites}, \textit{\ssr}, \textit{139}, 399--436,
  \doi{10.1007/s11214-008-9413-5}, 2008.

\bibitem[{\textit{{Tarduno} et~al.}(2010)}]{tarduno:2010}
{Tarduno}, J.~A., et~al., {Geodynamo, Solar Wind, and Magnetopause 3.4 to 3.45
  Billion Years Ago}, \textit{Science}, \textit{327}, 1238--,
  \doi{10.1126/science.1183445}, 2010.

\bibitem[{\textit{{Wilde} et~al.}(2001)\textit{{Wilde}, {Valley}, {Peck}, and
  {Graham}}}]{wilde:2001}
{Wilde}, S.~A., J.~W. {Valley}, W.~H. {Peck}, and C.~M. {Graham}, {Evidence
  from detrital zircons for the existence of continental crust and oceans on
  the Earth 4.4Gyr ago}, \textit{\nat}, \textit{409}, 175--178, 2001.

\bibitem[{\textit{Williams}(2007)}]{williams:2007}
Williams, Q., {Water, the Solid Earth, and the Atmosphere: The Genesis and
  Effects of a Wet Surface on a Mostly Dry Planet}, in \textit{Treatise on
  Geophysics,Volume 9}, pp. 121--143, Elsevier, 2007.

\end{thebibliography}

\end{document}